\documentclass[5p]{elsarticle}

\usepackage{lineno,hyperref}
\modulolinenumbers[5]

\journal{Journal of \LaTeX\ Templates}

\biboptions{numbers,sort&compress}
\usepackage[status=draft,layout=footnote]{fixme}
\usepackage{amsmath}
\usepackage{graphicx}
\graphicspath{}

\usepackage{color,soul}
\newcommand{\rev}[1]{\textcolor{black}{#1}}

\usepackage{pdfpages}

\begin{document}
\setcounter{page}{0}
\thispagestyle{empty}
\begin{figure*}
	\centering
	\includegraphics[width=\textwidth]{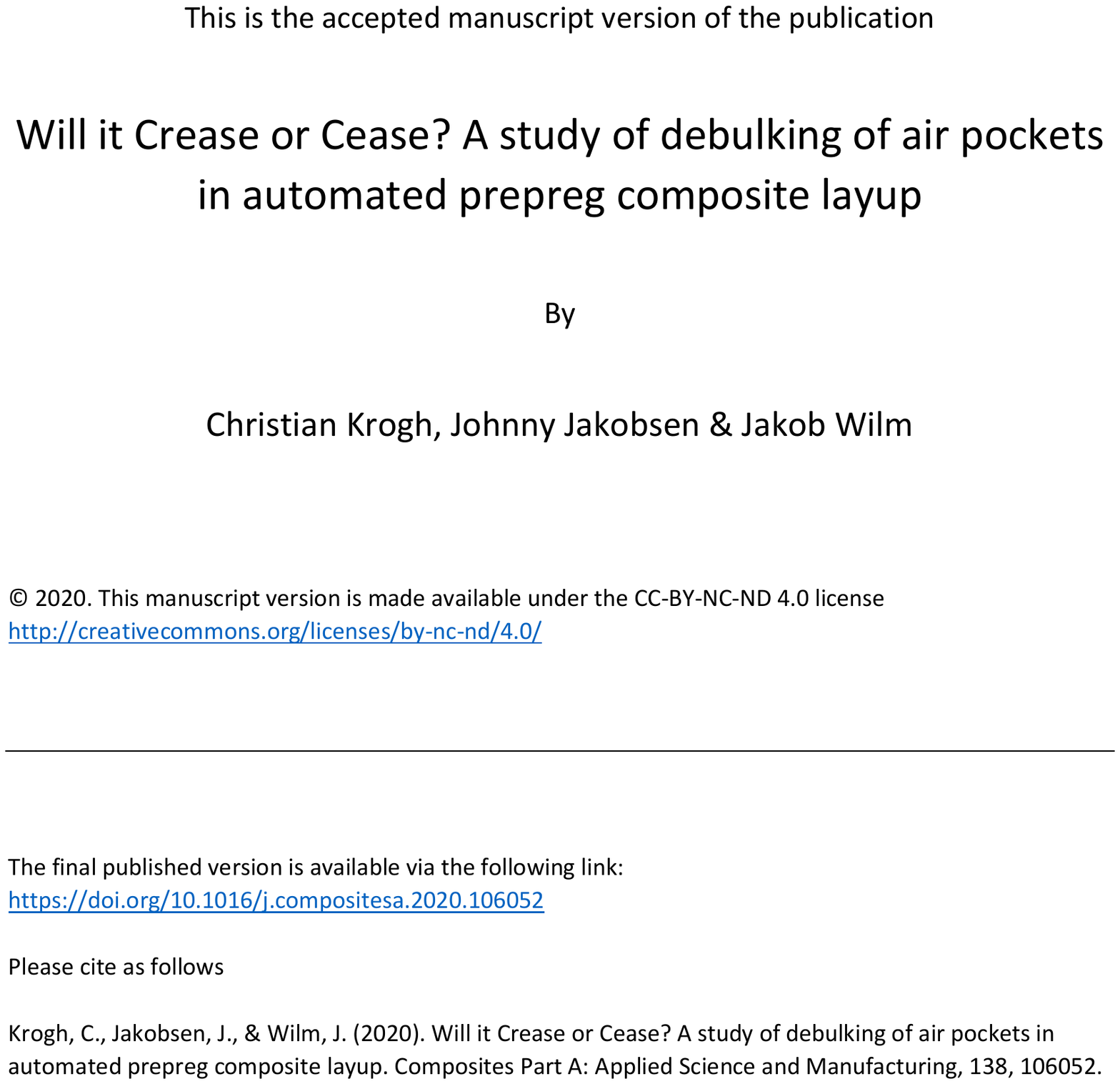}
\end{figure*}

\begin{frontmatter}

\title{Will it Crease or Cease\rev{?} A study of Debulking of Air Pockets in Automated Prepreg Composite Layup}

\author{Christian Krogh\corref{mycorrespondingauthor}}
\cortext[mycorrespondingauthor]{Corresponding author}
\ead{ck@mp.aau.dk}
\author{Johnny Jakobsen}
\address{Department of Materials and Production, Aalborg University, Fibigerstraede 16, 9220 Aalborg, Denmark.}

\author{Jakob Wilm}
\address{The Maersk Mc-Kinney Møller Institute, University of Southern Denmark, Campusvej 55, 5230 Odense M, Denmark.}

\begin{abstract}
Automatic draping of carbon-fiber prepreg plies for the aerospace industry is a promising technique for lowering the manufacturing costs and to this end, a thorough in-process quality control is crucial. In this paper, out-of-plane defects in the layup are investigated. After draping, air pockets are occasionally encountered. The question is, if such apparent defects can be mitigated sufficiently during vacuum debulking. The 3D topology is measured by means of a structured-light 3D scanner and air pockets are segmented. An approximate mass-spring ply model is used to study the behavior of the air pockets during application of vacuum pressure. The model is computationally fast and will indicate online whether the air pocket will be removed or manual intervention is required. Upon comparing the model predictions with experimental data, it is shown that the system is capable of correctly predicting 13 out of 14 air pockets in a test layup.
\end{abstract}


\end{frontmatter}


\section{Introduction}
Composites are known for their excellent mechanical properties but suffer from high manufacturing costs. One cause of the expensive manufacturing is the involvement of manual labor, which is especially pronounced for small batch production. A particularly costly step is the manual \textit{draping} or lay-up of the fiber plies in the mold. To this end, the FlexDraper research project \cite{Ellekilde2020, Gunnarsson2018} seeks to develop an automatic ply layup solution for the aerospace industry which has the potential of decreasing the manufacturing costs and at the same time increasing the part quality. The robot system has the ability to manipulate and drape entire plies of woven prepreg fabric onto double-curved molds of low curvature. A vision system is implemented for generating feedback to the controller.   

\begin{figure}[tb]
	\centering
	\includegraphics[width=1.0\linewidth]{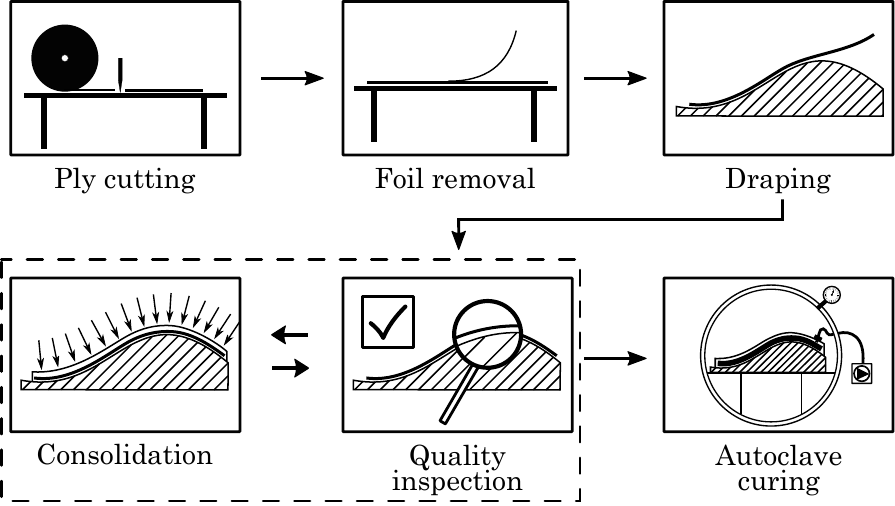}
	\caption{Major steps in the production of prepreg-based composite parts. All steps except ``Autoclave curing'' are automated in the FlexDraper projekt.}
	\label{fig:prepregprocess}
\end{figure}

The manufacturing of prepreg-based composite parts in the aerospace industry consists of a number of steps prior and subsequent to the draping of the plies as sketched in Figure~\ref{fig:prepregprocess}. Prior steps include for instance ply cutting and backing foil removal. Subsequent to draping, each ply is consolidated, e.g. by means of vacuum debulking, such that the plies are compacted and any entrapped air is removed. Next, the quality of the draped ply must be inspected based on a number of parameters. One very important requirement is close conformity of the draped ply to the mold surface. With the robot system, this conformity is checked by recording the ply surface using an optical 3D measurement system. The consolidated plies are, however, impractical to remove from the layup in case the requirement of mold-ply conformity is not met. For this reason, each ply is also recorded before consolidation. \rev{In this context, an area where the ply does not conform to the mold surface within a certain tolerance is denoted an \textit{air pocket}. The challenge is hence to identify air pockets and predict the outcome of the consolidation.} \textit{That is, will any air pockets formed during the automatic draping be sufficiently mitigated during consolidation?}

In an ideal automatic ply layup process no air pockets will form. There are, however, circumstances which are expected to influence the ply-mold conformity. Such circumstances include running-in of a new part, i.e. mold-ply system, but also statistical uncertainty on the material behavior as well as tolerances on the hardware and sensors are expected to cause minor air pockets. On the other hand, hardware defects or build-up of resin from the plies can cause major air pockets. In this case, the process has to be halted and the problem corrected.        

Numerous researchers have treated the challenge of developing process models applicable to fiber plies. The models are typically based on experimental testing and thus the mechanical behavior of the fiber plies has also been a topic of great interest. In general, a woven fabric has a high stiffness in the fiber directions while the fiber tows can rotate at the cross-over points, i.e. shear, thus enabling the deformation from flat to double-curved \cite{Cao2008}. The out-of-plane bending stiffness has also been identified as important - especially for the wrinkling behavior \cite{Boisse2011}. Prepreg plies, i.e. fabric that is pre-impregnated with a partially cured resin, further posses a tacky interface with the mold surface \cite{Newell1995}. 

General draping process models are often accomplished using the finite element (FE) method due to its versatility and the opportunities of advanced material and interface model definitions, see e.g. \cite{Boisse2011, Peng2005, Harrison2016, Krogh2019}. Also approximate models have been developed, for instance kinematic mapping methods \cite{Mack1956}, discrete mass-spring models or particle models \cite{Breen1994, Boubaker2006} or a finite-difference solution to a large-deflection plate equation \cite{Do2006}. 

More specific consolidation models have also been reported. An advanced predictive FE consolidation model was applied to UD prepregs in the study by \cite{Belnoue2018}. The model is hyper-viscoelastic and accounts for both squeezing flows and bleeding flows. For the former, the fibers and resin deform together in an incompressible manner. For the latter, only the resin flows between the fibers and the compaction is limited and described by a laminate bulk factor. The \rev{results} showed that the main drivers for wrinkle formation are the bulk factor, the incompressibility of prepreg material in the direction of the fibers, and the resistance of the plies to slip with respect to another due to the friction or the boundary conditions. The outcome is an excess length which is accommodated through in-plane or out-of-plane buckling.

In the study by \cite{Thompson2018} the consolidation of layers of dry textiles was modeled using a FE model developed for draping simulation. In the \rev{model}, the authors modified the pressure-overclosure relationship used in the penalty contact formulation between the layers. In this way, the transverse compressibility can be taken into account. The debulking process is simulated with a uniformly applied pressure to the top layer. The method was demonstrated on an external radius of a C-section geometry, which led to the development of wrinkles.

\rev{A continuum consolidation model was developed by \cite{Simacek2020} and implemented using a semi-implicit finite difference scheme. The model accounts for fiber deformation, resin flow and porosity. The numerical examples provided insight into important physical effects of the process which the authors concluded must be elaborated with further experimental work.}

Simpler, analytical models have also been developed. In the work by \cite{Dodwell2014}, a model for consolidation of an arbitrary number of plies on an external radius with flanges on either side was presented. If the friction of the flanges can resist sliding, the fibers on the radius will be constrained and wrinkling can develop. Otherwise, if sliding of on the flanges occur, so-called \textit{book-ends} can form. Wrinkling is considered a stability problem. The model performed reasonably in predicting the buckling wavelength in comparison to an experiment with consolidation on a C-section. 

An analytical model for consolidation of laminates on mold sections of constant corner radius has also been presented \cite{Levy2019}. It includes a convex, a concave and a U-shaped corner. The authors introduced a \textit{conformation number}, $\Lambda$, which depends on the radius, flange length and coefficient of friction. Depending on the value of $\Lambda$ the compaction can either be friction or pressure dominated. Friction dominated entails that the frictional resistance of the plies will prevent sliding and thus the plies will not conform to the mold. Pressure dominated entails that the plies can slide, but that the curvature of the mold can result in different consolidation pressures across the thickness of the laminate. Apart from the quantities in $\Lambda$, also the bulk factor of the material enters into the model expression. Using a semi-empirical rule of mixtures between friction and pressure dominated behavior, the model was shown to predict thickness variations well for a high number of experiments.

From the previously mentioned references it can be concluded that parameters such as the mold curvature, frictional resistance, laminate thickness and bulk factor are important for the outcome of the consolidation process. \rev{Similar conclusions have also been reported based on experimental work \cite{Hallander2013}}. In the work by \cite{Lightfoot2013}, wrinkles were hypothesized to arise from shear stresses due to a mismatch between the coefficient of thermal expansion for the mold and the plies. The authors also investigated the effect of using a release film between the mold and the first ply. This was found to facilitate slipping of the ply stack which in turn leads to wrinkling of the first ply in a corner radius. The effect of debulking after every fourth ply in stead of at the end of the layup did not improve the quality significantly. In-plane waviness was also observed and is likewise related to the developed shear stresses.

The studies listed above all concern layups with multiple plies that are assumed to be closely conforming to the mold but where the mold curvature can cause wrinkling during compaction, e.g. as seen with an external radius of a C-section. For application with the FlexDraper robot system, a model must be able evaluate the air pockets of a single ply. The evaluation must be carried out while the robot system is in operation, i.e. \textit{online}. The FE-based approaches have the versatility to model this behavior but are too time consuming. For this reason, an approximate mass-spring model is applied which can still take the material properties and frictional resistance of the prepreg plies into account. Because the model only concerns a single ply, the thickness and bulk factor are not a concern. 

The rest of the paper is organized as follows: In Section \ref{sec:robotcell} the automatic draping and debulking process is briefly elaborated. In Section \ref{sec:meth} the process of ply scanning, air pocket segmentation, meshing and debulk analysis is explained. Section \ref{sec:results} presents experimental results of a draped ply and its debulking along with the model predictions. Section \ref{sec:discussion} discusses the applicability of the results and possible improvements to the model. Lastly, a conclusion completes the paper.    


\section{Automatic Draping and Debulking} \label{sec:robotcell}
\begin{figure}[tb]
	\centering
	\includegraphics[width=\linewidth,trim={0 0 0 0},clip]{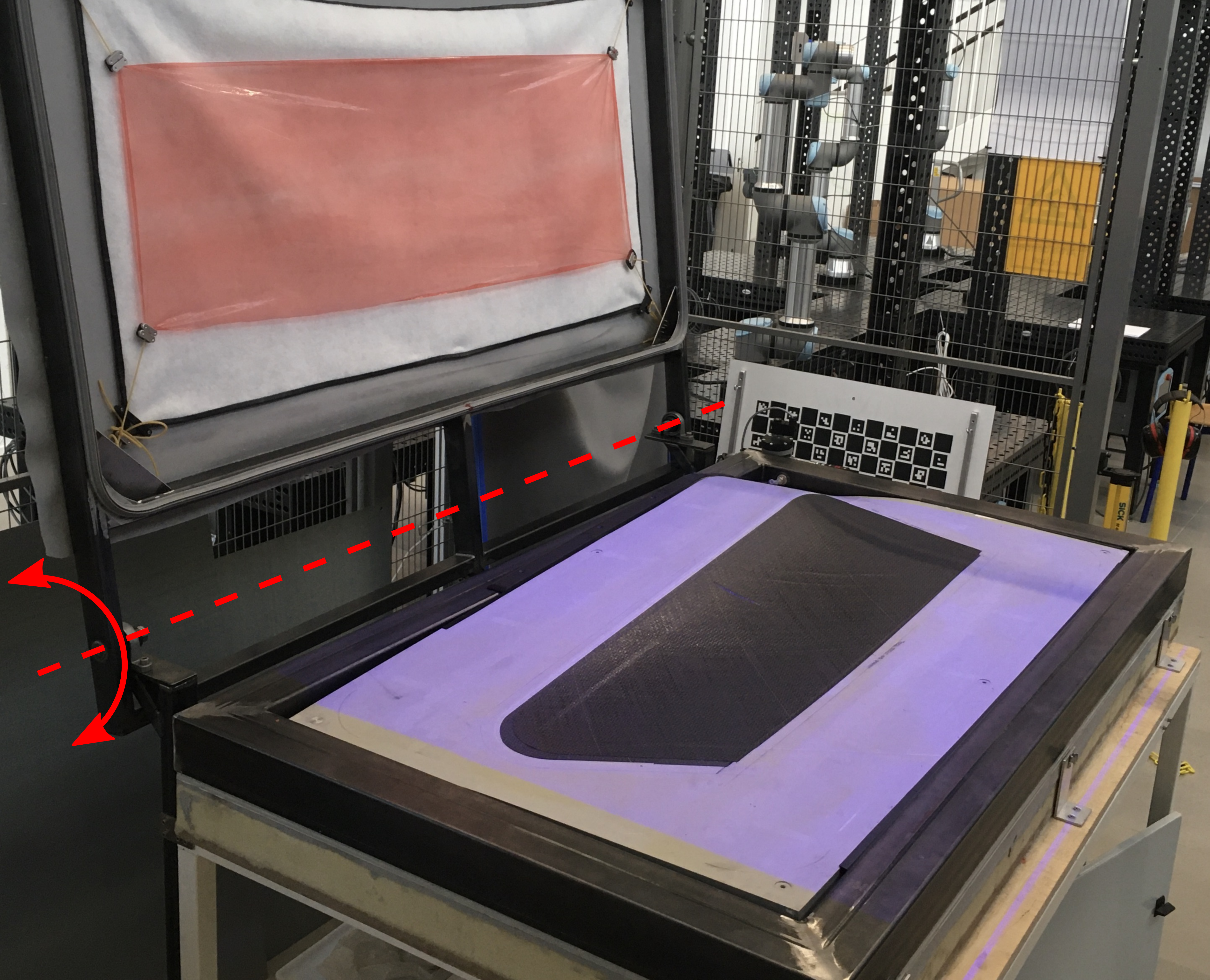}
	\caption{Mold with draped ply and automatic debulking system. Blue projector light is seen shining on the surface.}
	\label{fig:debulk_cell}
\end{figure}

An overview of the FlexDraper system can be found in the paper by Ellekilde et al \cite{Ellekilde2020}. The core of the robot cell is a 6-axis industrial robot with a specially designed end effector consisting of a grid of actuated suction cups. The system is able to remove the backing foil from cut plies, drape the plies on the mold, consolidate the plies by means of vacuum debulking while monitoring the process using the vision system. A mold with the automatic debulking system is shown in Figure~\ref{fig:debulk_cell}. The debullking system consists of a steel frame mounted on the perimeter of the mold. A lid with an equivalent steel frame and a rubber membrane can be closed with a pneumatic cylinder thereby forming a sealed chamber around the mold. A breather fabric and a solid release film is mounted on the rubber membrane. Vacuum is achieved by means of a vacuum pump.   

\begin{figure}[tb]
	\centering
	\includegraphics[width=0.7\linewidth]{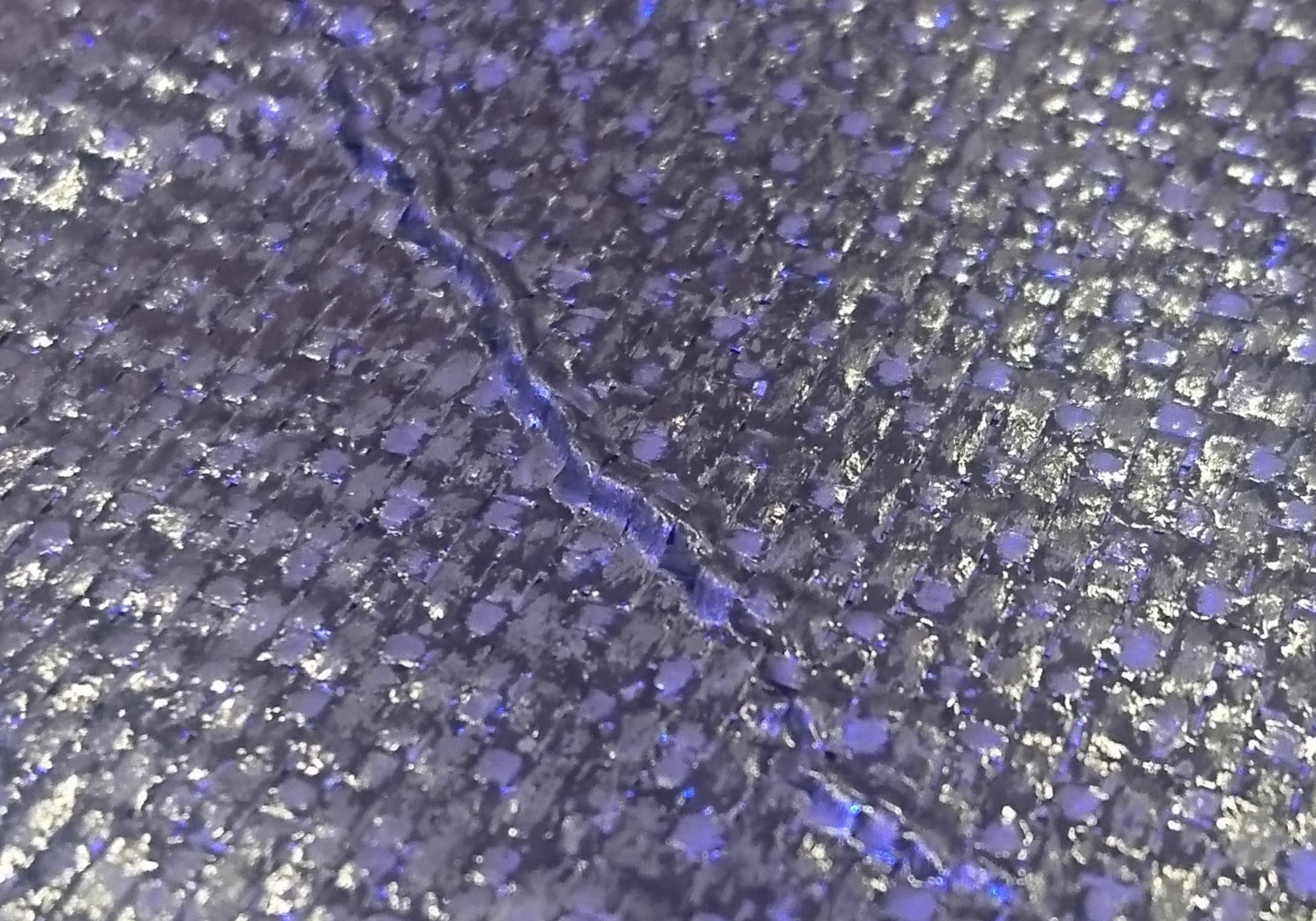}
	\caption{Typical wrinkle formed after debulking of air pocket. A wrinkle like this in the final part would necessitate scrapping of the entire part.}
	\label{fig:wrinkle}
\end{figure}

A demonstrator part in the shape of a generic aerospace composite component was chosen and is used throughout this study. A ply from this part can be seen draped on the mold in Figure~\ref{fig:debulk_cell}. As it can be seen, the part has a fairly low curvature but it has still presented itself as a challenge for the automatic robot system. The typical outcome of a debulking process where an air pocket is not mitigated, is presented in Figure~\ref{fig:wrinkle}. On that account, one approach could be to conservatively discard all draped plies containing air pockets. It has, however, been observed that some air pockets are mitigated completely and for this reason it is worth investigating the debulking process further.

\section{Methodology} \label{sec:meth}
The workflow of the air pocket evaluation starts with data acquisition, i.e. 3D scanning of the ply after draping whereby a point cloud is obtained. Next, possible air pockets are segmented by comparison of the 3D point cloud to a reference configuration. Air pockets detected in the layup are then exported for meshing and subsequent evaluation with the mass-spring ply model. Lastly, the results of the mass-spring model are post-processed with the intention of improved accuracy and independence of the discretization.

\subsection{Ply Scanning}
\begin{figure}
	\centering
	\includegraphics[width=\linewidth]{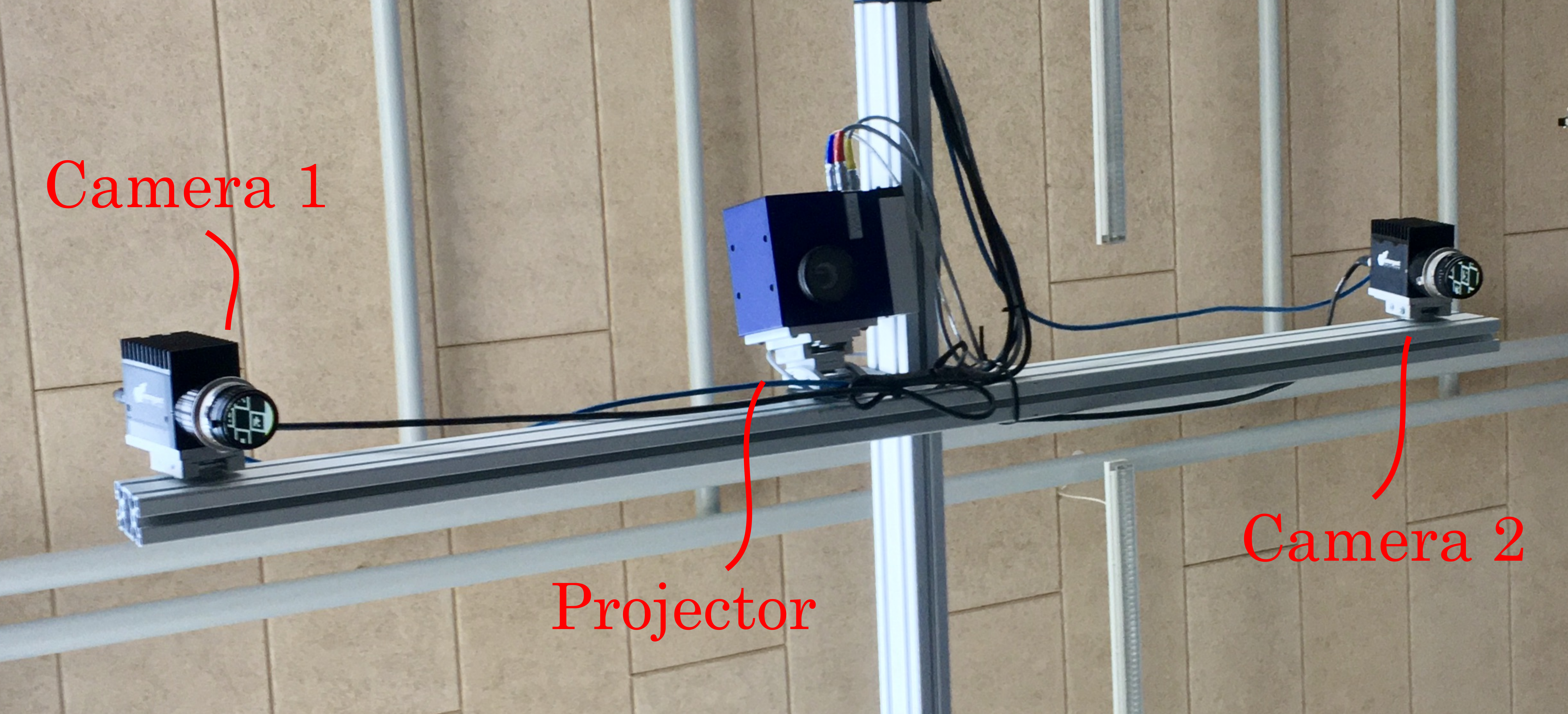}
	\caption{3D scanner setup mounted 2.4 m over the mold area. The view direction is vertically downward. The baseline between cameras is approximately 1 m.}
	\label{fig:IMG_3040}
\end{figure}

The measurement of the layup topology is performed using a custom developed high-fidelity 3D scanner. The scanner hardware consists of two 20MPx industrial cameras (EVT HT-20000) and a pattern projector (Vialux STAR-07) as shown in Figure \ref{fig:IMG_3040}. Every camera pixel, and hence triangulated surface point, corresponds to a surface area of approximately 0.3 mm$^2$. The camera lenses are equipped with narrow band-pass filters matching the projector's blue LED wavelength (460 nm). This arrangement provides good robustness against ambient illumination because the band-pass spectrum is heavily dominated by the projector's light.

This system was specifically developed to provide highly detailed scans of the layup with a minimum of noise. The surface reconstruction is based on the principle of structured light, whereby a series of patterns are  projected and captured synchronously by both cameras. The patterns are designed to encode points on the scan surface uniquely, which allows for determining point correspondences between the cameras. Based on these, depth is then recovered by means of triangulation.

While many coding strategies exist for structured light, this work is based on two frequency phase shifting fringe patterns which obey the heterodyne principle \cite{Reich1997}. In this technique, a number of sinusoidal patterns (sometimes referred to as fringe patterns) are projected onto the object surface, each with a fraction of a $2 \pi$ phase shift. For this work, 6 shifted patterns with 50 sine periods over the field of view (FOV) are used, then 6 shifted patterns with 51 sine wave periods, as experimentation has shown that this is a good trade-off between acquisition and reconstruction time, memory consumption and permissible phase noise. This technique is able to densely encode the surface using the phase information, leading to one triangulated 3D point per non-occluded camera pixel. 

Structured light surface encoding is dependent on light reflecting from the material surface. In this regard, the CFPR prepreg material presents itself as particularly challenging as it has a very wide reflectivity distribution. The fibers are very light absorbing from most viewing angles, while the resin has a high degree of gloss/specularity. To mitigate this, an HDR (high dynamic range) scanning sequence is adopted, whereby the pattern sequence and camera shutter levels are stepped through a series of exponentially growing exposure times. These HDR sequences are combined at the phase map level, where the phase is recovered at each pixel from the brightest non-saturated exposure level. This increases the effective dynamic range of the sensor manyfold and enables scanning with a minimum of noise. The scanned ply data is stored as an organized point cloud, i.e. with one point per camera pixel. 

\subsection{Segmentation} \label{sub:segmentation}
\begin{figure}[tb]
	\centering
	\includegraphics[width=0.8\linewidth]{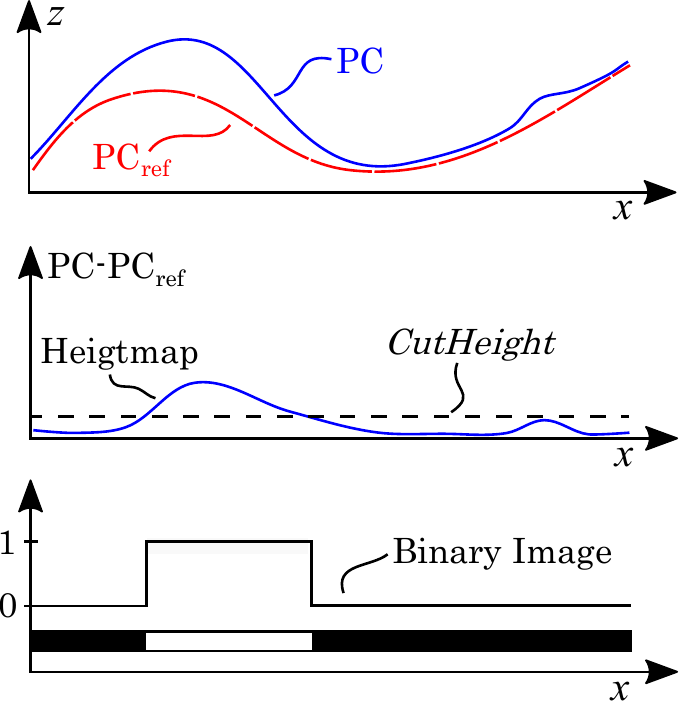}
	\caption{Segmentation of air pockets. From the top: Point cloud and reference point cloud, heightmap and cutting tolerance, and binary image of air pockets.}
	\label{fig:segmentation}
\end{figure}

The point cloud representing the draped ply surface is imported in MATLAB \cite{matlab}. Here, the point cloud is transformed to the mold coordinate system and noise is removed using the \textit{pcdenoise} function. The methodology is to set a threshold equal to one standard deviation from the mean of the average distance to neighbors of all points. If the average distance to its 4-nearest neighbors is above the threshold, the point is removed, see e.g. \cite{Rusu2008}. The surface is then filtered using a median filter. Next, using a reference surface recorded before draping of the current ply, a heightmap is generated, i.e. gridded XY-data with the difference between the current ply and the previous. This and the following operations are sketched in 2D in Figure~\ref{fig:segmentation}.

The heightmap surface is the basis for the segmentation of air pockets. The approach is to apply a fixed tolerance (\textit{CutHeight}) on the heightmap which has been empirically determined. Geometrically the operation corresponds to cutting the heightmap with a plane parallel to the XY-plane offset with the tolerance. Any points of the heightmap below the plane are set to 0 (black) while points above the plane are set to 1 (white). In this way a binary image of the air pockets is created. 

The binary image enables the use of image processing techniques to efficiently locate the air pockets in the heightmap. First the MATLAB function \textit{imfill} is called to fill any holes in the binary image, i.e. black pixels inside the regions of white pixels. This operation is based on morphological reconstruction \cite{Soille1999}. Next, \textit{bwboundaries} is called to locate the perimeter of the air pockets, i.e. regions of white pixels. The function is an implementation of the Moore-Neighbor tracing algorithm modified by Jacob's stopping criteria \cite{gonzalez2004}. 

For each air pocket, the enclosing ply surface and the reference surface is extracted. To facilitate the meshing (described in the next section), a portion of the reference surface outside the perimeter of the air pocket must also be extracted. The combination of these surfaces are referred to as a \textit{patch}.
\subsection{Meshing}
The objective of the meshing is to generate the initial coordinates of the nodes, i.e. masses, and their connectivity for the mass-spring model. The inputs to the meshing routine are the patch surfaces obtained from the segmentation, the \rev{local fiber angles in the center of the patches} and the discretization, i.e. the chosen distance between the nodes along the fiber directions. In this study, the same discretization is used for both fiber directions. The methodology of the meshing follows the pin-jointed net assumption which is also the basis for the kinematic mapping algorithms \cite{Mack1956}: the fiber stiffness is infinite while the shear stiffness is zero. 

First, two initial fiber paths emanating from the center of the patch are laid out on the patch surface in the directions of the \rev{input local} fiber angles. The initial fiber paths serve as constraints when calculating the coordinates of the nodes. Nodes can now be placed on the initial fiber paths such that the spacing corresponds to the discretization. The remaining nodes are placed - one by one - on the patch surface in the regions spanned by the initial fiber paths such that the distance to adjacent nodes corresponds to the discretization. This operation generates the net of pin-jointed nodes. Alongside with the node placement, the connectivity is stored. Thus, each node will connect with up to four other nodes. 
\begin{figure}[tb]
	\centering
	\includegraphics[width=\linewidth]{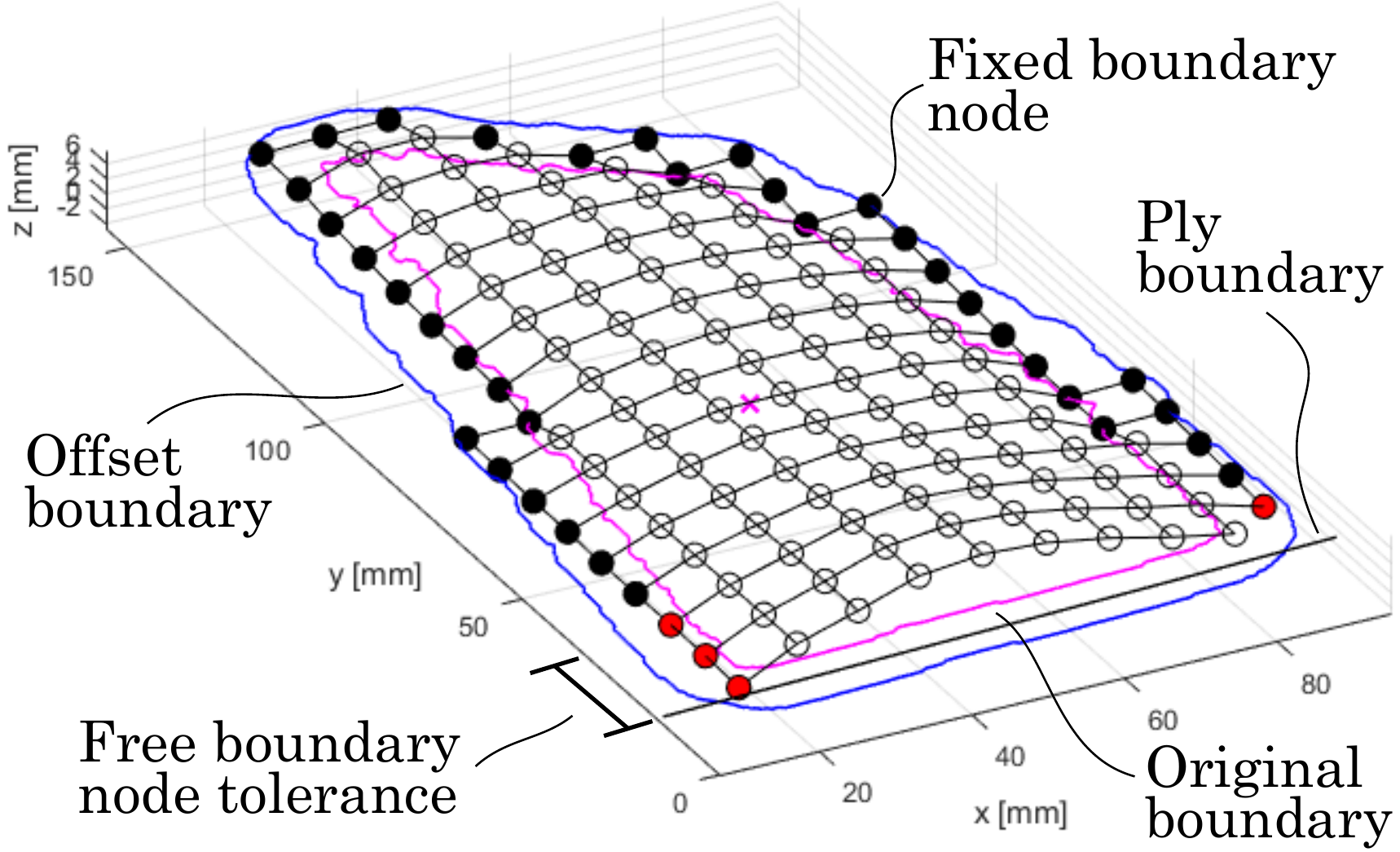}
	\caption{\rev{Generation of boundary nodes using an offset patch boundary. Black nodes are fixed boundary nodes while red nodes are free boundary nodes.}}
	\label{fig:boundary_nodes}
\end{figure}

To be able to use a rather coarse discretization, the boundary of the air pocket requires special attention: The original air pocket boundary is offset by a distance equal to the discretization. Next, the mesh of nodes is trimmed using the offset boundary. Nodes that are located between the original boundary and the offset boundary and are within a tolerance from the mold surface are now designated as \textit{boundary nodes}. The operation of trimming and locating of boundary nodes is sketched in Figure~\ref{fig:boundary_nodes}.

The behavior of the boundary nodes is chosen based on empirical observations during testing of the debulking system: The ply material surrounding the air pockets tends to constrain the movement of the original air pocket boundary. The exception to this is when the air pockets are located close to the boundary of the ply. Here, movement seems to be very likely. For this reason, the boundary nodes are split into two categories. The first category is \textit{free boundary nodes}, i.e. boundary nodes that are located within a tolerance of the ply boundary. The second category is \textit{fixed boundary nodes} whose coordinates will not change during the analysis. The distinction between free and fixed boundary nodes is sketched in Figure~\ref{fig:boundary_nodes}. The tolerance is set to 20 mm.

\subsection{Debulk Analysis}
With the mesh of nodes in the initial configuration, defined in the form of coordinates and their connectivity, the objective is to find the final configuration after application of vacuum pressure. To this end, the mass-spring modeling framework described by Ben Boubaker et al. \cite{Boubaker2006} is used as the starting point. The idea is to introduce springs at each node which account for the various deformation modes. The equilibrium configuration is found by minimizing the total potential energy of the system subject to kinematic constraints, i.e boundary conditions and mold contact. The authors of \cite{Boubaker2006} solved the nonlinear optimization problem analytically, but here it is solved numerically using a gradient-based method.

The elastic deformation modes accounted for in the present model include out-of-plane bending and in-plane shear. The fiber direction stiffness is several orders of magnitude higher than for shear and bending. For this reason, and due to the numerical condition of the system, it was decided to kinematically constrain the node-node distance along the fiber directions to the value of the discretization. Further, inequality constraints ensure that no penetrations of the reference surface occur.

\begin{figure}[tb]
	\centering
	\includegraphics[width=1.0\linewidth]{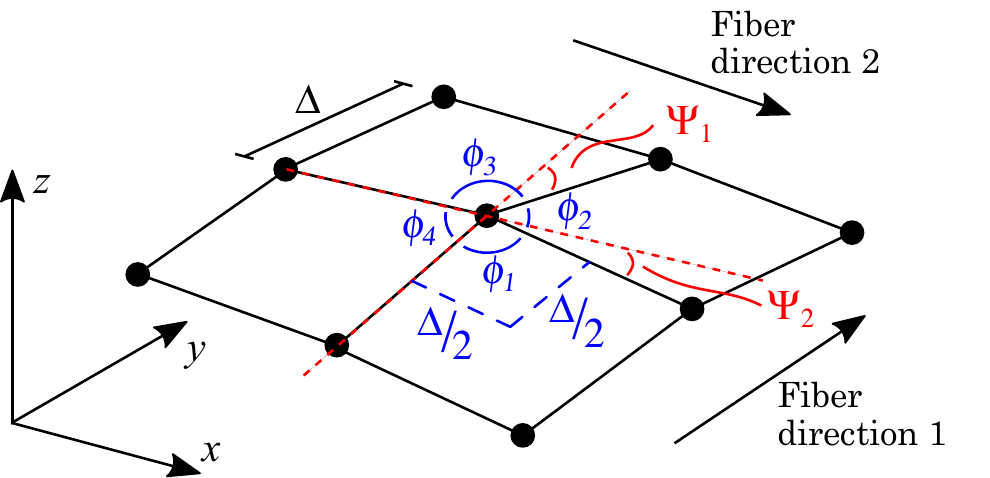}
	\caption{Deformation modes of mass-spring ply model.}
	\label{fig:plyenergy}
\end{figure}

Regarding the elastic deformations, their nodal energy contributions are defined in the following. The bending energy, $U_{bend}$ can be derived from the bending energy stored in a linear elastic beam subjected to pure bending \cite{Gere2008}:
\begin{align} 
	U_{beam} = \frac{EI \, \theta^2}{2 L}
\end{align}
In the formula, $EI$ is the flexural rigidity, $\theta$ is the angle of the bent arc shape and $L$ is the length of the beam. Using the second moment of area, $I$, for a rectangular cross section and defining the length and the width of the beam to be equal to the discretization $\Delta$, the bending energy for the $i$th node is obtained:   
\begin{align}
	U_{\mathrm{bend},i} = \frac{1}{2} \frac{E \, t^3}{12} \bigg[ (2\Psi_1)^2 + (2\Psi_2)^2\bigg]
\end{align}
Here, $\Psi_1$ and $\Psi_2$ are the out-of-plane bending angles for each fiber direction as sketched in Figure~\ref{fig:plyenergy}. From trigonometry, $\Psi$ is equal to one half of $\theta$. The ply thickness is denoted by $t$. 

The shear energy can be derived by considering the work done by a shear stress $\tau$ during shearing of a cubic material element by the angle $\gamma$. Assuming linear elastic behavior, the strain energy density is \cite{Gere2008}:
\begin{align}
	u_{shear} = \frac{1}{2} \, \tau \, \gamma
\end{align} 
Using Hooke's law, i.e. $\tau = G \gamma$ with $G$ being the shear modulus, and multiplying by the volume, the $i$th nodal shear energy, $U_{\mathrm{shear},i}$, becomes:
\begin{align}
	U_{\mathrm{shear},i} &= \frac{1}{2}\, G \left(\frac{\Delta}{2}\right)^2 t \, \left(\gamma_1^2 + \gamma_2^2 + \gamma_3^2 + \gamma_4^2 \right) \\
	~~ &, \gamma_j = \mathrm{abs}\left(^\pi{\mskip -5mu/\mskip -3mu}_2 - \phi_j\right) , j=1,...,4  \notag
\end{align}
Here $\gamma_1$ - $\gamma_4$ are the four shear angles that can be calculated for each node by using the angles $\phi_1$ - $\phi_4$ from Figure~\ref{fig:plyenergy}. The volume is defined from the thickness, $t$, and a square with side lengths $\Delta/2$ and represents the volume sheared by a single $\gamma$, i.e. the area enclosed by the blue dashed line in Figure~\ref{fig:plyenergy}. 

The ply is subjected to two external loads: gravity and the applied vacuum pressure. The energy associated with gravity on the $i$th node is defined as follows:
\begin{align}
	U_{\mathrm{grav},i} = \frac{A_\mathrm{patch,ply} \, t \, \rho}{N} \, g \, \delta_{z,i}
\end{align} 
Where $A_\mathrm{patch,ply}$ is the surface area of the ply patch, $\rho$ is the mass density of the ply and $N$ is the number of nodes. In this way, the fraction corresponds to the nodal mass. Further, $g$ is the gravitational constant and $\delta_{z,i}$ is the movement in the $z$-direction of the node relative to the initial configuration.

Pressure is a quantity that acts perpendicular on a surface. With the flexible membrane of the automatic debulking system facilitating the vacuum, it is, however, not certain that the load is transferred in a normal direction to the ply. For this reason, and due to the relatively flat molds in the project, the vacuum pressure in this study will act in the $z$-direction:
\begin{align}
	U_{\mathrm{vac},i} = \frac{P \, A_\mathrm{patch,ply}}{N} \, \delta_{z,i}
\end{align}  
In the formula, $P$ is the magnitude of the vacuum pressure (1 bar). The fraction corresponds to the equivalent nodal force.

Next, the free boundary nodes are considered. The movement of a free boundary node is accompanied by a frictional energy contribution:
\begin{align}
	U_{\mathrm{fric},i} = \delta_i \, \mu \, \frac{P \, A_\mathrm{patch,ply}}{N} 
\end{align}   
Here, $\delta_i$ is the movement of the node relative to the initial configuration, and $\mu$ is the coefficient of friction.

The total potential energy of the system of masses and springs is then the sum of internal nodal energies minus the external nodal energies:
\begin{align}
	\Pi &= \Pi_\mathrm{int} - \Pi_\mathrm{ext} \\
	\Downarrow~& \notag\\
	\Pi &= \sum_{i=1}^{N} U_{\mathrm{bend},i} + U_{\mathrm{shear},i} + U_{\mathrm{fric},i} - U_{\mathrm{grav},i} - U_{\mathrm{vac},i}
\end{align}
The minimum of the total potential energy $\Pi$ is found, as previously mentioned, using a gradient-based optimization algorithm. The optimization problem is formulated as follows:
\begin{align}
	\underset{\mathbf{d}}{\mathrm{minimize}}&~~ \Pi(\mathbf{X}) \qquad, \mathbf{X} = \mathbf{X_{ini}} + \mathbf{d}\notag\\
	\text{s.t.}&~~ F_\mathrm{ref}(x_{j},y_{j}) - z_{j} \leq 0 \notag \\
	 &~~ ||\{x_k, y_k, z_k\} - \{x_l, y_l, z_l\}|| - \Delta = 0 \\
	 &~~\forall j ~,~j=1,...,N \notag\\
	 &~~\forall k,l ~,~k \in \mathbf{C_1}  \quad , l \in \mathbf{C_2} \notag
\end{align}
In the expression, the design variables, $\mathbf{d}$, are the displacements relative to the initial nodal coordinates $\mathbf{X_{ini}}$. The sum of the former and the latter is thus the current nodal coordinates, i.e. $\mathbf{X} =$ $[x_1, y_1, z_1,...,x_N,y_N,z_N]$. The first family of constraints (inequality) ensures that no mold penetrations occur using a function for the reference surface, $F_\mathrm{ref}(x,y)$. The second family of constraints (equality) enforces the node-node distance to be equal to the discretization $\Delta$. The bookkeeping, i.e. unique node connectivity, is stored in the vectors $\mathbf{C_1}$ and $\mathbf{C_2}$.

The formulation of the objective function and the constraints enables the implementation of analytical gradients. The exception is the first constraint with the function $F_\mathrm{ref}(x,y)$, i.e. an implicit function of the design variables. Here, semi-analytical gradients are implemented.


\subsection{Post-Processing the Results} \label{sub:pp}
The results obtained with the mass-spring model will naturally be dependent on the discretization used. By lowering the discretization sufficiently, it is believed that the solution will converge. However, wrinkles encountered with the debulking system (Figure~\ref{fig:wrinkle}) are typically less than a millimeter in height. A sufficient discretization which would be able to capture such wrinkles adequately would then entail very many nodes and thus infeasible computation times. For this reason, it was decided to develop a post-processing approach such that the mass-spring model can be evaluated with a reasonably coarse discretization while the end result ideally will not be sensitive to the discretization.

\begin{figure}[tb]
	\centering
	\includegraphics[width=0.8\linewidth]{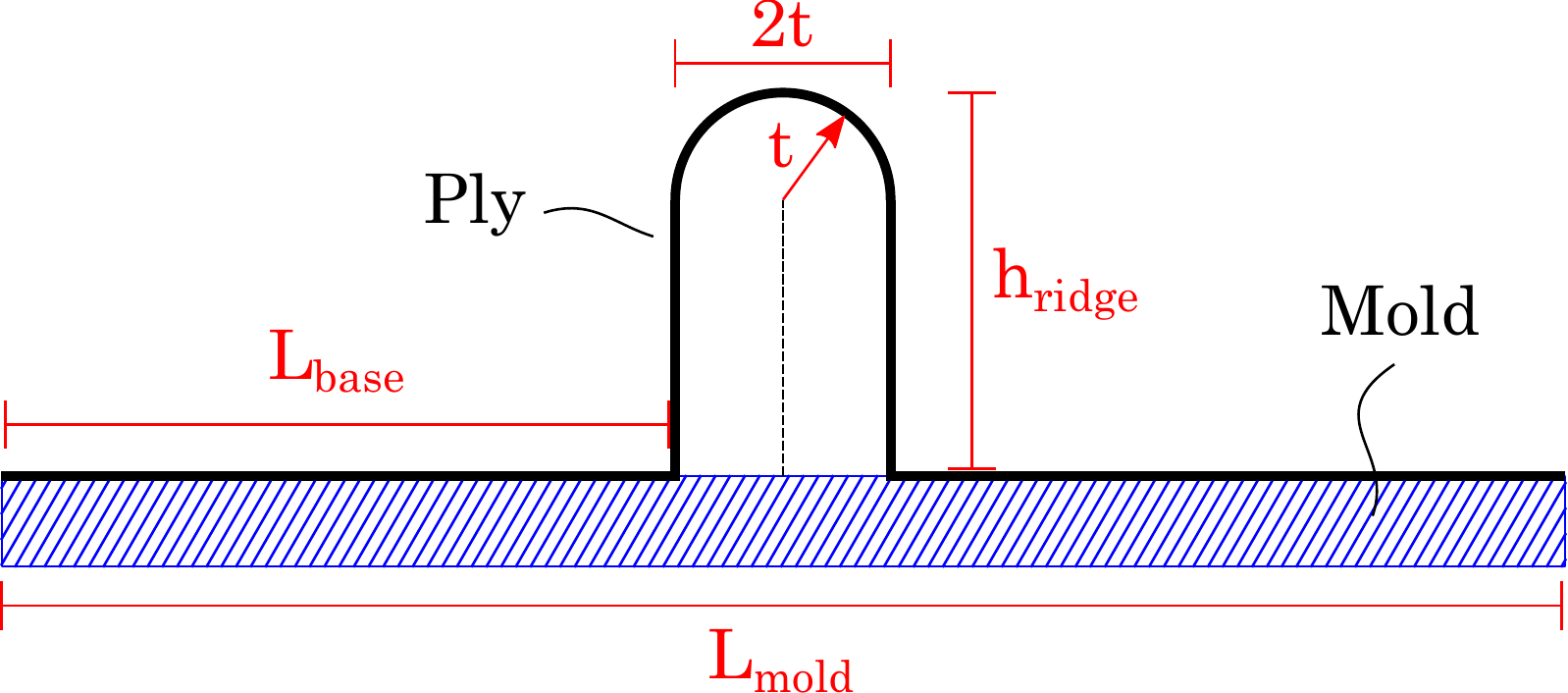}
	\caption{The geometrical wrinkle model for post-processing results.}
	\label{fig:ppmodel}
\end{figure}

As it was seen in Figure~\ref{fig:wrinkle}, the wrinkles formed typically take the shape of a ridge. In this regard, the formation of a ridge is simplified to a 2D problem, i.e. a determination of the cross-sectional width and height. The post-processing model is sketched in Figure~\ref{fig:ppmodel}, and the cross-sectional height of the ridge will be obtained solely based on geometrical parameters.

The assumption is that width of the ridge cross-section will be equivalent to two ply thicknesses. Physically, this is the smallest width that is possible to achieve without self intersections. Additionally, the top of the ridge cross-section is assumed to be a semi-circle with a radius equal to the ply thickness. The height of the ridge is determined based on the excess ply length compared to the mold arc length, $L_\mathrm{mold}$. Thus, the first step is to calculate the amount of the ply length that can fit on the mold:
\begin{align}
L_\mathrm{base} = \frac{L_\mathrm{mold} - 2 t}{2}
\end{align}  
Next, the excess ply length is used to determine the height of the ridge cross-section:
\begin{align}
h_\mathrm{ridge} = \frac{L_\mathrm{ply} - 2 L_\mathrm{base} - \pi t}{2} + t
\end{align} 
In the expression, $L_\mathrm{ply}$ is the ply length.
\begin{figure}[tb]
	\centering
	\includegraphics[width=1.0\linewidth]{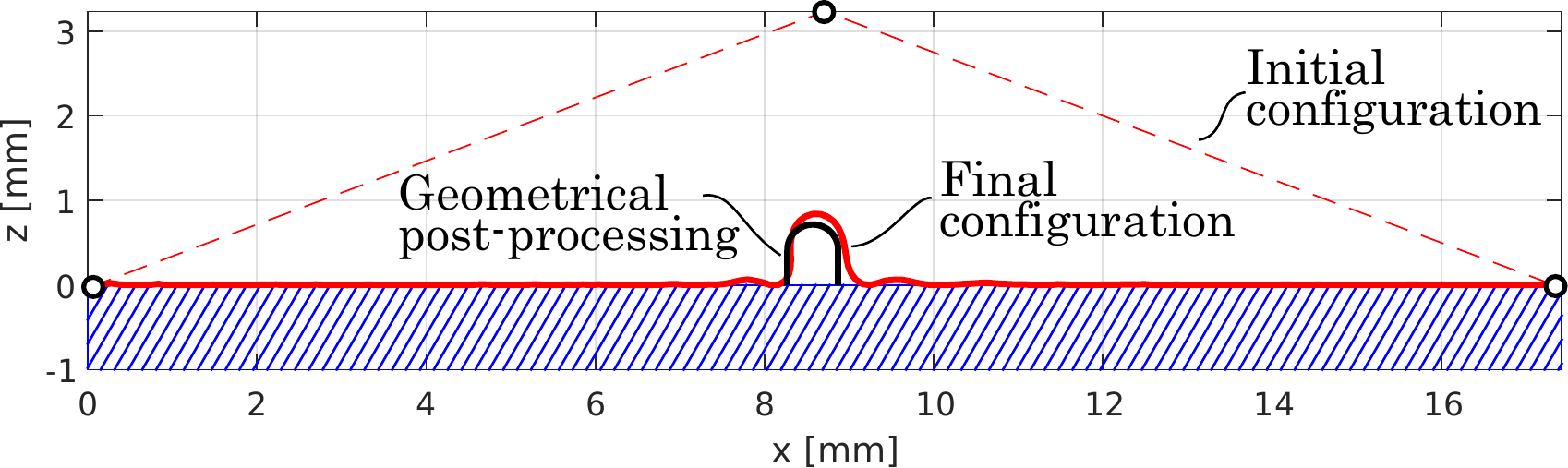}
	\caption{Formation of a wrinkle in 2D model. The red ply is debulked onto the blue hatched mold. The geometrical post-processing is shown in black.}
	\label{fig:2dwrinklep2}
\end{figure}

The idea is to apply the geometrical post-processing model after a solution with the mass-spring model has been obtained. As previously discussed, a rather coarse discretization will give a poor wrinkle representation and such a situation is exemplified in 2D in Figure~\ref{fig:2dwrinklep2}. Here, three nodes connected by red dashed lines are seen of which two (left and right) are on the mold whereas one (center) is located 3.2 mm above the mold. With the chosen discretization of 9.3 mm, this situation will be the final wrinkle representation of the mass-spring model. To investigate the effect of a fine discretization, the red dashed line connecting the nodes has been discretized into 200 segments. This 2D model is now evaluated analogously to the 3D model presented in the previous section with the exception that shear is not included. Further, instead of solving for the initial configuration in one step, the vacuum pressure (1 bar) is split up in five load steps. Before each load step, the direction of the vacuum is updated such that it is perpendicular on the ply surface. 

\begin{figure}[t]
	\centering
	\includegraphics[width=1.0\linewidth]{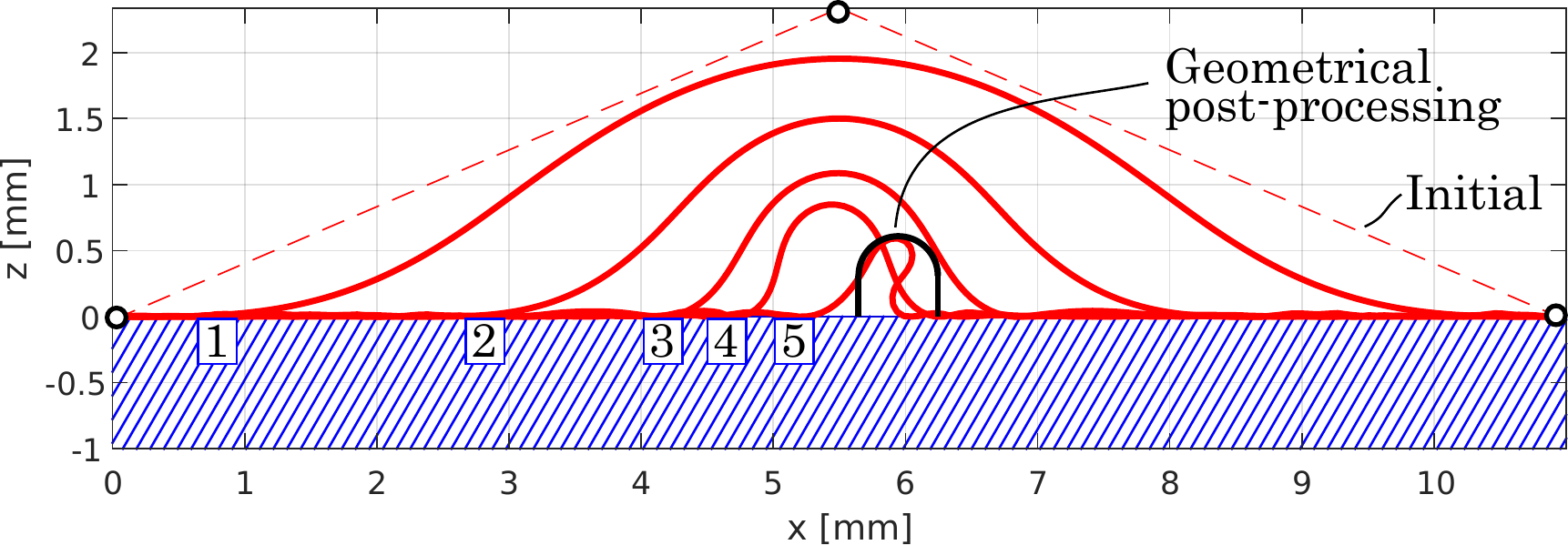}
	\caption{Evolution of a wrinkle in 2D model during five load steps (numbers in blue rectangles). The red ply is debulked onto the blue hatched mold. The geometrical post-processing is shown in black.}
	\label{fig:2dwrinklep7}
\end{figure}

The final configuration of the ply in Figure~\ref{fig:2dwrinklep2} does indeed resemble the cross-section of a ridge and it also agrees well with the post-processing model. Another example is presented in Figure~\ref{fig:2dwrinklep7} in which it is also possible to see the evolution of the wrinkle in each load step. In this latter example the discretization is 6.0 mm. Again, a good agreement between the 2D ply model and the geometrical post-processing model is seen. 

A final remark about the post-processing model is that the assumption of a wrinkle width equal to two times the ply thickness only seems to hold true for certain small wrinkle heights. That is, if the wrinkle becomes higher it also tends to become wider. However, when a wrinkle reaches a certain size, it will result in an infeasible layup and it will thus have to be discarded anyway. In that sense, the assumption on the wrinkle width makes the predictions conservative.   
\begin{figure*}[tb]
	\centering
	\includegraphics[width=1.0\linewidth]{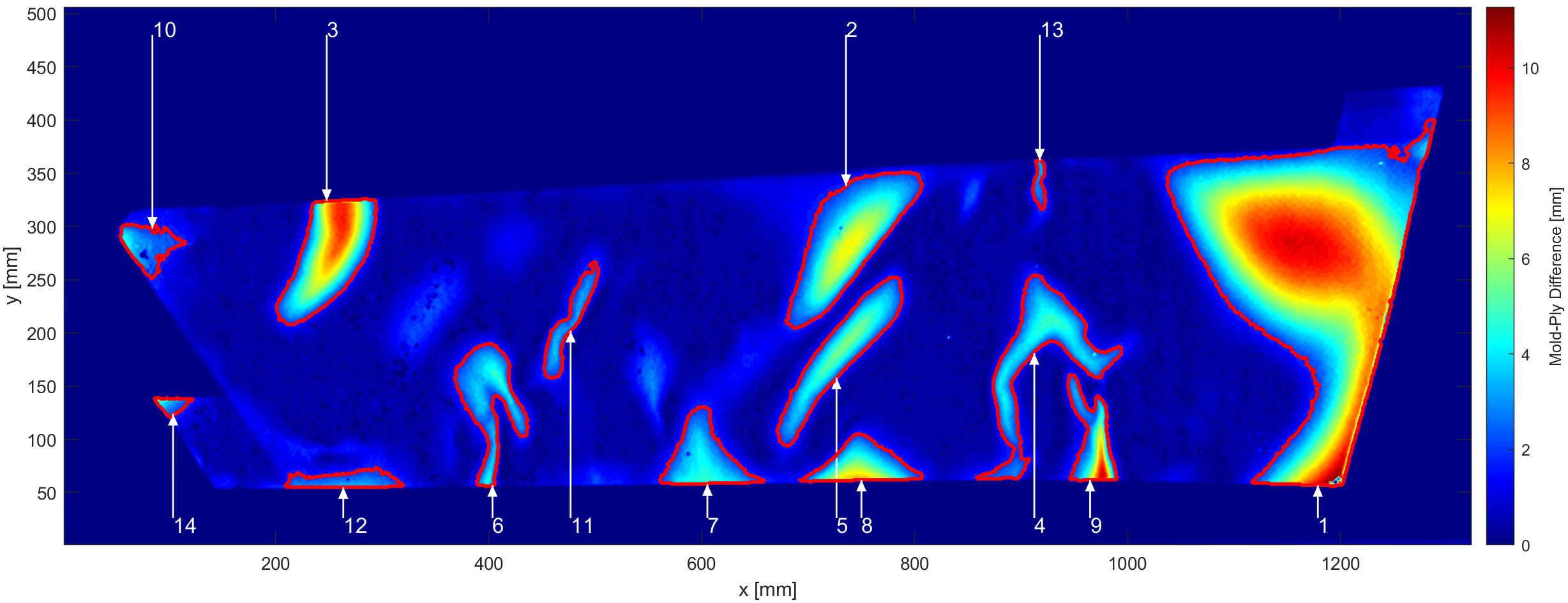}
	\caption{Segmented air pockets in layup. The order of the numbers is based on the air pocket sizes.}
	\label{fig:visiondrape}
\end{figure*}

Concerning the implementation, the approach is to start with the node with the largest ply-mold difference. From this node, the ridge cross-section defined by nodes is identified, the geometrical model is employed and the ply-mold difference is updated. These steps are continued with the next node, i.e. with the second-largest ply-mold difference etc. Throughout the post-pro-cessing, only nodes that have not previously been adjusted can be adjusted. Finally, after post-processing the entire mesh, all nodes - except nodes defining the ridges - are offset from the mold by the consolidated thickness $t/\beta$, i.e. with $t$ being the nominal ply thickness and $\beta$ the bulk factor. This operation will account for the ply thickness and thereby make the predictions after the geometrical post-processing comparable to the experimental results. The bulk factor has not been measured for the present material but instead the consolidated thickness has been determined from scans of debulked plies to 0.25 mm, i.e. such that $\beta = 1.2$. This value agrees well with a value of 1.23 reported in the literature for a similar material \cite{Levy2019}.      


\section{Results} \label{sec:results}
In this section, results with the debulk analysis framework is presented. The basis is the demonstrator part in Figure~\ref{fig:wrinkle}. First, experimental drape data are obtained. Next, air pockets are segmented and evaluated with the model. Lastly, the model predictions are compared to the result of debulking the draped ply.

\subsection{Draping}
For data generation with the demonstrator part, a ply was manually draped onto the mold. It was carried out in a manner such that the ply was aligned with respect to the prescribed boundary but with a number of air pockets of varying sizes and heights. The layup was then scanned with the optical 3D measurement system and processed to segment the air pockets. The result is presented in Figure~\ref{fig:visiondrape}. The settings used for the segmentation are presented in Table~\ref{tab:seg_set}. The tolerance used to cut the heightmap (\textit{CutHeight}) was introduced in Section~\ref{sub:segmentation}. Further, two additional tolerances are included to discard insignificant air pockets below a certain size which could for instance arise from noise in the data. These tolerances are \textit{AreaTol} and \textit{PeakTol}, which are lower limits on patch surface area and peak, respectively.
\begin{table}[tb]
	\centering
	\caption{Settings used for air pocket segmentation.}
	\begin{tabular}{ccc}
		\noalign{\smallskip}\hline\noalign{\smallskip}
		\textit{CutHeight} & \textit{AreaTol} & \textit{PeakTol}\\
		\noalign{\smallskip}\hline\noalign{\smallskip}
		2.0 mm & 2.0 cm$^4$ & 1.25 mm \\
		\noalign{\smallskip}\hline
	\end{tabular}
	\label{tab:seg_set}
\end{table}

From Figure~\ref{fig:visiondrape} it is seen that 14 air pockets were detected in the layup. Their surface areas range from 5.35 cm$^2$ (air pocket \#14) to 438 cm$^2$ (air pocket \#1). Some are located in the interior of the ply, e.g. air pocket \#5 while others are located such that their boundary is also the ply boundary, e.g. air pocket \#9.

\subsection{Air Pocket Evaluation}
The patch surfaces were then exported for meshing. Regarding the discretization it was chosen to target a specific number of nodes (100) for all patches rather than using a fixed distance specification between nodes. The reason is that the air pockets tend to have a bell-shaped surface regardless of their size which can thus be represented well by a size-independent mesh. Further, this approach has the advantage that the computational time is more well-defined and not air pocket size dependent.   

The mass-spring model was then evaluated for each air pocket. The parameters used for the model are listed in Table~\ref{tab:mat_par}. The prepreg material used in this study exhibits nonlinear and rate-dependent behavior as was shown in previous publications \cite{Krogh2019, Krogh2019c}. The representation of, respectively, the bending and shear behavior through the constant moduli $E$ and $G$ deserve some elaboration. Concerning the bending behavior, the value of Young's modulus, $E$, was found by using the static deflection from a cantilever test \cite{Krogh2017a}. Concerning the shear behavior, the value of the shear modulus, $G$, was found from the initial behavior recorded through bias-extension and picture-frame tests \cite{Krogh2019, Krogh2019c} and also from considerations on providing sufficient resistance against crumpling of the mesh. The coefficient of friction $\mu$ was measured in connection with a previous study \cite{Krogh2019}.
\begin{table}[tb]
	\centering
	\caption{Parameters used for mass-spring model.}
	\begin{tabular}{ccccc}
		\noalign{\smallskip}\hline\noalign{\smallskip}
		 t & E & G & $\rho$ & $\mu$ \\
		\noalign{\smallskip}\hline\noalign{\smallskip}
		 0.3 mm & $3.6 \, 10^8$ Pa & $4.0 \, 10^7$ Pa & 1048 kg/m$^3$ & 0.4 \\  
		\noalign{\smallskip}\hline
	\end{tabular}
	\label{tab:mat_par}
\end{table}
  
\begin{figure}[tb]
	\centering
	\includegraphics[width=1.0\linewidth]{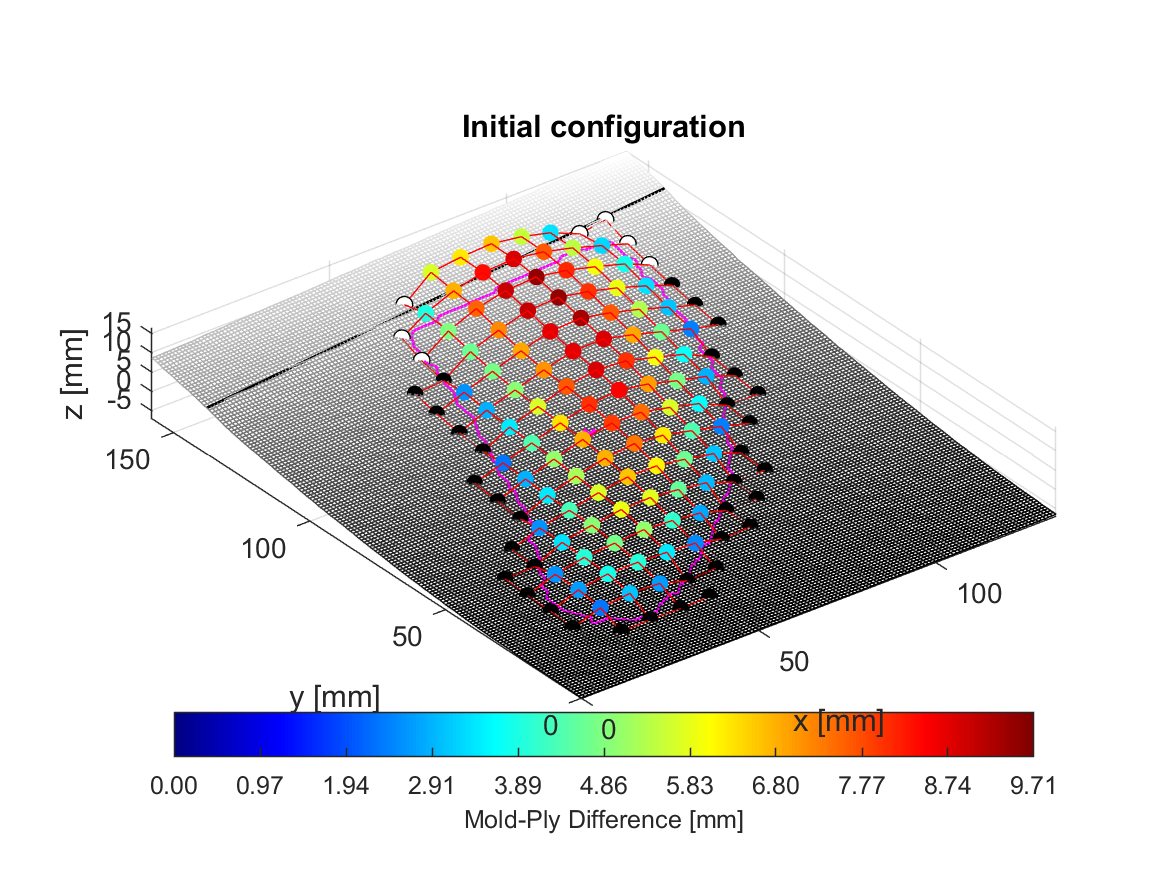}
	\includegraphics[width=1.0\linewidth]{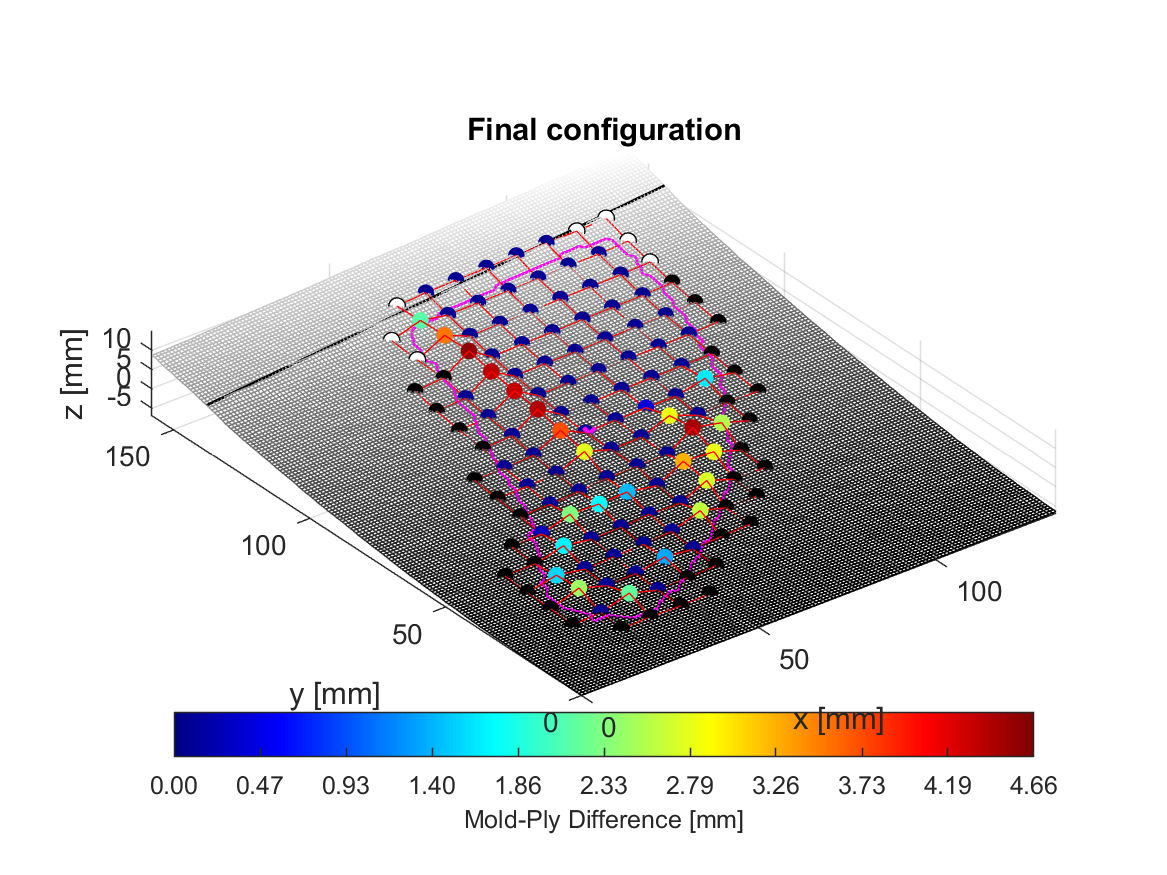}
	\caption{Evaluation of air pocket \#3: Initial configuration after meshing (top) and final configuration after application of vacuum (bottom). The black nodes are fixed boundary nodes while the white nodes are free boundary nodes.}
	\label{fig:patch3ini}
\end{figure}

An example of an air pocket being evaluated with the mass-spring model is provided in Figure~\ref{fig:patch3ini}. In the top part of the figure, the patch is seen after meshing and is thus in the initial configuration. In the bottom part of the figure, the final configuration has been found after application of vacuum. In the figure, it is seen that the model predicts a wrinkle in the form of a ridge. The maximum height of the ridge above the mold is 4.66 mm. However, as previously discussed this number is sensitive to the discretization. Upon using the geometrical post-processing approach, the maximum value reduces to 1.57 mm. Recall the over-predictions of the post-processing approach discussed in Section~\ref{sub:pp}. Still, the experimental result from patch \#3 is a ridge with a height of 1.40 mm (the experimental data for comparison is elaborated in the following).   
  
\subsection{Comparison to Experimental Debulk Data}
\begin{figure*}[tb]
	\centering
	\includegraphics[width=1.0\linewidth]{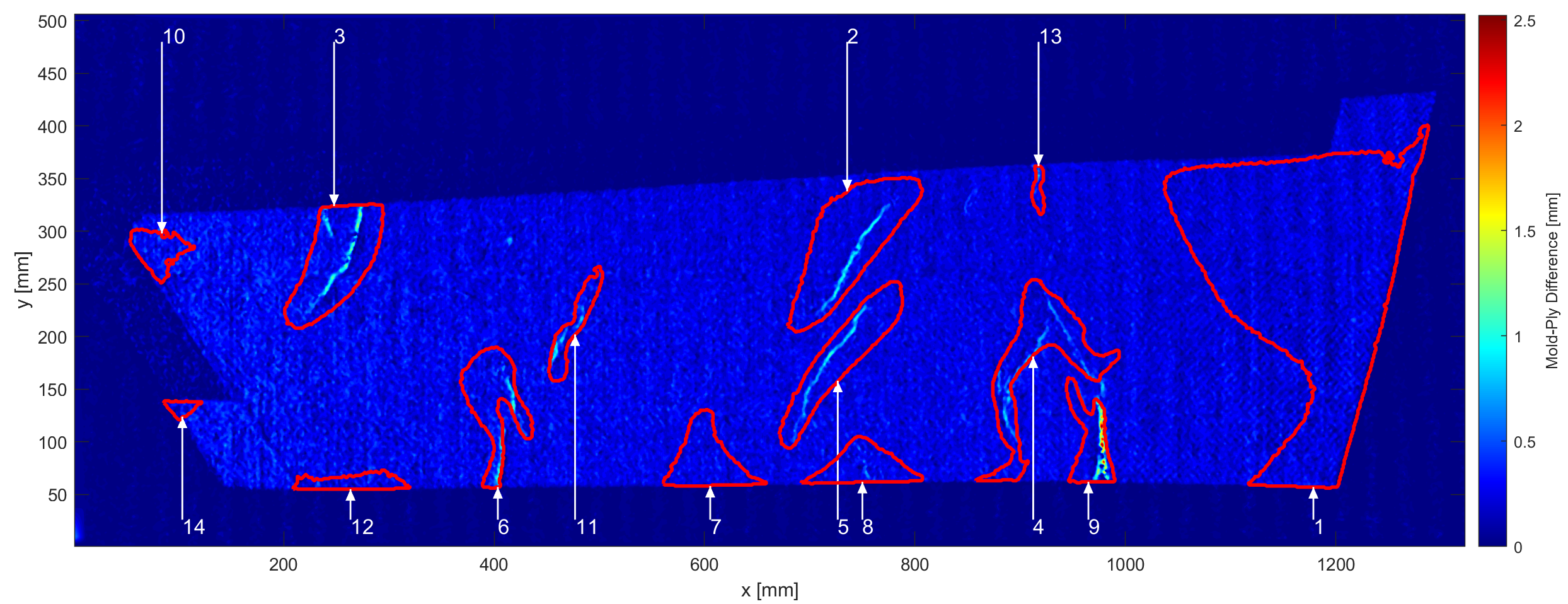}
	\caption{Heightmap of debulked ply with original air pocket contours overlayed.}
	\label{fig:visiondebulk}
\end{figure*}

To compare the model predictions with the experimental results of the debulked ply, it is convenient to use a single number. It was decided to use the root mean square (RMS) error of the ply-mold differences for this comparison. Regarding the experimental results, a heightmap of the debulked ply is generated with the original air pocket contours overlayed as presented in Figure~\ref{fig:visiondebulk}. The data points inside these contours, is the basis for the RMS calculation.
         
\begin{figure}[tb]
	\centering
	\includegraphics[width=1.0\linewidth]{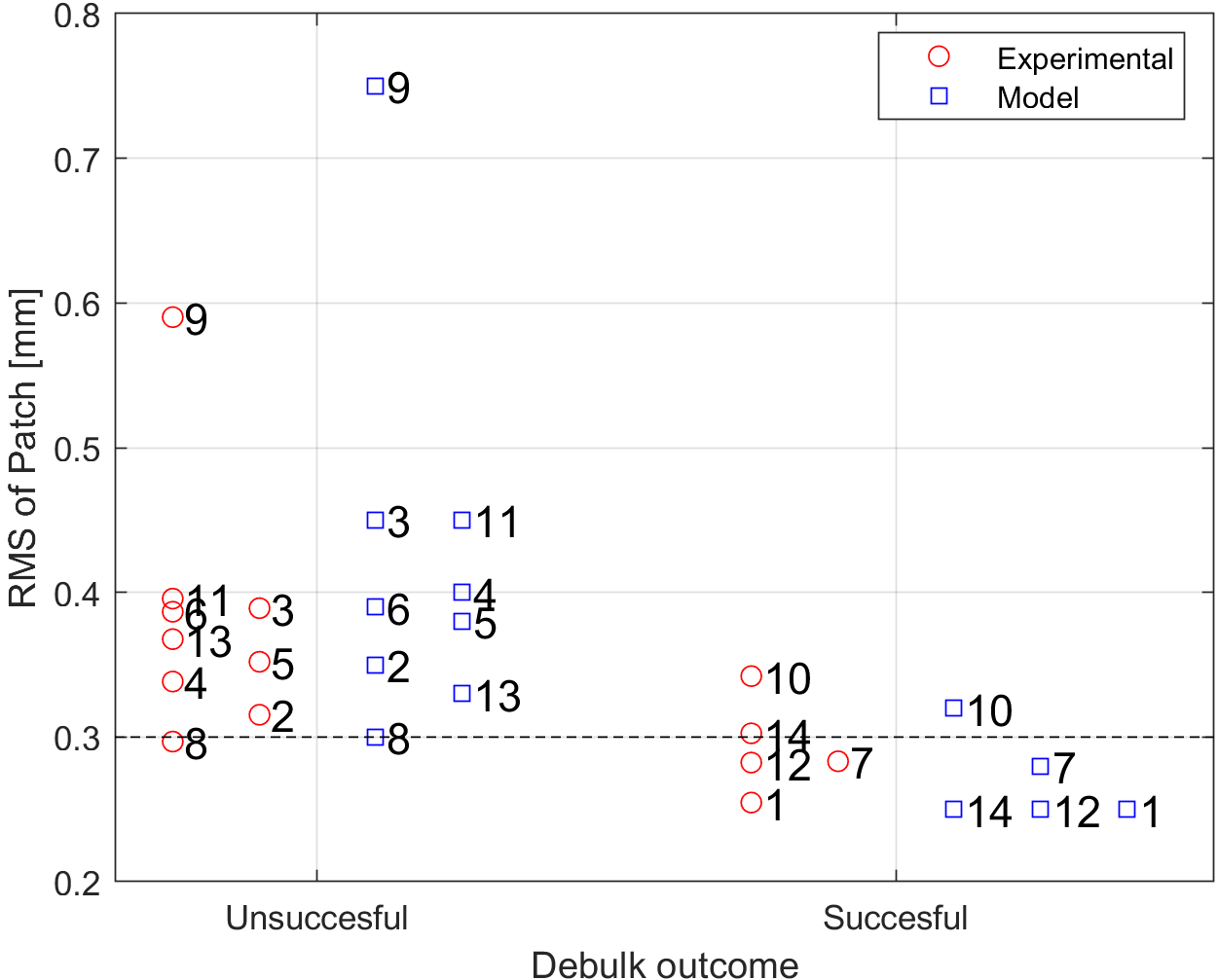}
	\caption{Comparison of RMS values from model predictions and 3D scan of debulked ply.}
	\label{fig:rmscomparison}
\end{figure}

Figure~\ref{fig:rmscomparison} presents the results of the RMS comparison between the model with post-processing and the experimental results. The debulking of each air pocket has been categorized as either \textit{unsuccessful} or \textit{successful} depending on whether there is a visible defect in the layup or not. A horizontal line (black dashed) has been drawn at 0.3 mm, i.e. the ply thickness. This line could serve as a distinction between unsuccessfully and successfully debulked air pockets. In this way, successfully debulked air pockets should have a RMS value between the consolidated ply thickness (0.25 mm) and the nominal ply thickness (0.3 mm). 

From Figure~\ref{fig:rmscomparison} it is evident that nine air pockets have an unsuccessful outcome while five air pockets have a successful outcome. For the successful air pockets of the experiment, \#10 and \#14 have a RMS above the divider line whereas the remaining three patches are below the line. The reason for the inconsistency is that the left part of the recorded point cloud has a high degree of noise which thereby increases the RMS for those patches. Air pocket \#1 which is also mitigated completely has a RMS of 0.25 mm.              
  
Concerning the unsuccessful air pockets of the experiment, all are above the divider line, except air pocket \#8, which has a RMS of 0.297. In the debulked ply, a very small wrinkle is visible and thus it was categorized as unsuccessful. A more advanced criterion than the RMS value could maybe help to discard this air pocket. The topic of a discard criterion is revisited in the discussion (Section~\ref{sec:discussion}).
  
Looking at the model predictions, it is seen that there is a reasonable agreement with the debulked ply data. For patches with no to small wrinkles, e.g. \#1, \#7, \#8, the agreement is excellent. The prediction for air pocket \#14 in fact matches better with what can be observed on the debulked ply rather than the scanned data. Air pocket \#10 suffers from noise also in the point cloud recorded before debulking (Figure~\ref{fig:visiondrape}) which in turn affects the meshing of the surface. Thus, a greater effort must be put into obtaining data with less noise or maybe using more filtering. With larger wrinkle height, the discrepancy between the the model and the experiment increases as e.g. seen with air pockets \#3 and \#9. This discrepancy is, as previously discussed, likely an effect of the assumptions behind the geometrical post-processing. But, as also noted previously, the predictions are conservative. The model RMS results are also presented in Table~\ref{tab:mod_discr} along with the discretizations used and the solution times and function evaluations logged.     
\begin{table}[tb]
	\centering
	\caption{Discretizations, RMS, solution times and function evaluations for air pocket analysis. The solution time includes meshing, evaluation with the mass-spring model and post-processing and is benchmarked on a standard laptop.}
	\begin{tabular}{ccccc}
		\noalign{\smallskip}\hline\noalign{\smallskip}
		& Discr. & RMS & Sol. time & Func. eval. \\
		& [mm] & [mm] & [s] & [-] \\
		\noalign{\smallskip}\hline\noalign{\smallskip}  
		\#1 & 23.2 & 0.25 & 6.07 & 140 \\
		\#2 & 9.33 & 0.35 & 9.80 & 439 \\
		\#3 & 8.40 & 0.45 & 8.62 & 543 \\
		\#4 & 7.97 & 0.40 & 11.2 & 480 \\
		\#5 & 7.18 & 0.38 & 9.41 & 459 \\
		\#6 & 5.97 & 0.39 & 14.2 & 717 \\
		\#7 & 6.49 & 0.28 & 8.33 & 566 \\
		\#8 & 5.70 & 0.30 & 11.9 & 744 \\
		\#9 & 5.25 & 0.75 & 12.5 & 999 \\
		\#10 & 4.59 & 0.32 & 7.41& 661 \\
		\#11 & 3.89 & 0.45 & 47.5& 565 \\
		\#12 & 5.93 & 0.25 & 5.70& 586 \\
		\#13 & 2.57 & 0.33 & 14.4& 1564\\
		\#14 & 2.44 & 0.25 & 21.9& 1480 \\
		\noalign{\smallskip}\hline
	\end{tabular}
	\label{tab:mod_discr}
\end{table}   
\section{Discussion} \label{sec:discussion}
The previous section has presented some promising results. Based on the RMS value and a threshold equal to the nominal ply thickness, the model was able to predict whether debulking was successful or unsuccessful for all air pockets except one. For this exception, noise in the input data to the model is a plausible explanation for the discrepancy. To provide a more robust categorization, a more advanced criterion could be developed. One idea is to include the maximum value of the patch surface. Doing so, the experimental debulk results will, however, become sensitive to noise. Further, the wrinkle shape predicted by the model becomes more important, because the same excess ply length can develop into multiple wrinkle configurations.

The geometrical post-processing was developed by studying wrinkle development in a 2D ply model. Rather than post-processing the results of the mass-spring model, the model could maybe be evaluated iteratively with local mesh refinement in areas where wrinkles form. The computational performance vs. the accuracy would then have to be evaluated.  

The model is quick to evaluate with the average evaluation time for an air pocket on a standard laptop being 13.5 s. The frequency of air pockets occurring in a production setup is not known, but if many air pockets have to be checked, the model can be run in parallel. Further improvements on performance can for instance be achieved by examining the mesh before evaluation of the mass-spring model. If a fiber path begins and ends in a fixed boundary node and the length of the fiber path is greater than the projected mold arc length by a certain amount, the air pocket could maybe be categorized as unsuccessful already at this stage.    

\rev{The in-plane shear and out-of-plane bending of the mass-spring model represent some basic but important deformations. A possible extension of the model could be to also account for torsion/twisting as well as in-plane bending of the fiber tows. The latter deformation has recently been pointed out in connection with the so-called \textit{second gradient effects} to provide more realistic tow deformations \cite{DAgostino2015}. Such an extension of the model will require additional material data but it not expected to increase the computational cost significantly.}

\rev{The meshing routine must receive the local fiber angles of a patch as an input. These fiber angles can for instance be approximately predicted by a kinematic mapping analysis of the entire ply on the mold as employed in this work. Because the fiber angles of the draped plies are also subject to the quality inspection in the FlexDraper robot system, a current task in the project is to implement a sensor for recording the fiber angles. A natural extension of the present work would be to employ the recorded fiber angles as the input to the patch meshing. }

\section{Conclusion}
This paper has presented an approach for investigating the occurrence of out-of-plane defects in automated composite ply layup. Using a structured-light 3D scanner, a ply can be scanned after draping and subsequently after vacuum debulk. The idea is to apply a mass-spring model to predict the outcome of the vacuum debulk which will either enable the process to continue or necessitate manual intervention.

Air pockets are segmented based on the point cloud of the draped ply and image processing tools. The segmented air pocket surfaces are then meshed such that nodes/masses are distributed on the surface while obeying the fabric kinematics. The final configuration of the mass-spring model is found by minimizing the total potential energy using a gradient-based optimization algorithm. The energies in the potential arise from out-of-plane bending, in-plane shear, friction of movable boundary nodes, gravity and the vacuum pressure. Kinematic constraints enforce the node-node distance and prevent mold penetrations.

To favor computational performance such that the model can run online, a rather coarse discretization is used. To this end, a geometrical post-processing approach was developed to give predictions which are less sensitive to the chosen discretization. The post-processing relies on an assumption of a typical wrinkle shape and thickness observed on the automatic debulking system, and was developed based on a finely discretized 2D ply model. 

The point cloud of the debulked ply was used for comparison with the model predictions. Here, good agreement was found when comparing the root mean square (RMS) error of ply-mold differences. The best agreement was found with air pockets which were completely mitigated or left a small wrinkle after debulking. This corresponds well with the situations where the assumptions behind the geometrical post-processing were met adequately. All in all, the model was found to give conservative predictions in that it over-predicts larger wrinkles. For these defects, manual intervention is required anyway.

The computational time of the numerical air pocket analysis was also found to be favorable. With a more refined discard criterion in combination with more experimental evidence to establish a threshold value, the approach presented in this paper will become a valuable tool for the automated ply layup system. 

\section*{Acknowledgements}
Funding: The work presented in this paper is part of the research project ``FlexDraperProduct: Developing an integrated production-ready solution for automated layup of prepreg fiber plies'', funded by Innovation Fund Denmark (grant no. 8057-00011B). The support is gratefully acknowledged.

\section*{Data Availability}
The raw data required to reproduce these findings are available to download from \url{http://dx.doi.org/10.17632/mjtzzcym3g.1}.

\bibliography{./paper_library}

\begin{thebibliography}{10}
\expandafter\ifx\csname url\endcsname\relax
  \def\url#1{\texttt{#1}}\fi
\expandafter\ifx\csname urlprefix\endcsname\relax\def\urlprefix{URL }\fi
\expandafter\ifx\csname href\endcsname\relax
  \def\href#1#2{#2} \def\path#1{#1}\fi

\bibitem{Ellekilde2020}
L.-P. Ellekilde, J.~Wilm, O.~Nielsen, C.~Krogh, E.~Kristiansen, G.~Gunnarsson,
  T.~Stenvang, J.~Jakobsen, M.~Kristiansen, J.~Glud, M.~Hannemose,
  H.~Aan{\ae}s, J.~Kruijk, I.~Sveidahl, A.~Ikram, H.~Petersen, {Design of
  Automated Robotic System for Draping Prepreg Composite Fabrics}, Robotica
  (2020) 1--17.~In press.
\newblock \href {http://dx.doi.org/10.1017/S0263574720000193}
  {\path{doi:10.1017/S0263574720000193}}.

\bibitem{Gunnarsson2018}
G.~G. Gunnarsson, O.~W. Nielsen, C.~Schlette, H.~G. Petersen, {Fast and Simple
  Interacting Models of Drape Tool and Ply Material for Handling Free Hanging,
  Pre-impregnated Carbon Fibre Material}, in: G.~O., M.~K. (Eds.), Informatics
  in Control, Automation and Robotics. ICINCO 2018. Lecture Notes in Electrical
  Engineering, Vol. 613, Springer, Cham, 2020, pp. 1--25.
\newblock \href {http://dx.doi.org/10.1007/978-3-030-31993-9_1}
  {\path{doi:10.1007/978-3-030-31993-9_1}}.

\bibitem{Cao2008}
J.~Cao, R.~Akkerman, P.~Boisse, J.~Chen, H.~S. Cheng, E.~F. de~Graaf, J.~L.
  Gorczyca, P.~Harrison, G.~Hivet, J.~Launay, W.~Lee, L.~Liu, S.~V. Lomov,
  A.~Long, E.~de~Luycker, F.~Morestin, J.~Padvoiskis, X.~Peng, J.~A. Sherwood,
  T.~Stoilova, X.~Tao, I.~Verpoest, A.~Willems, J.~Wiggers, T.~Yu, B.~Zhu,
  {Characterization of mechanical behavior of woven fabrics: Experimental
  methods and benchmark results}, Composites Part A: Applied Science and
  Manufacturing 39~(6) (2008) 1037--1053.
\newblock \href {http://dx.doi.org/10.1016/j.compositesa.2008.02.016}
  {\path{doi:10.1016/j.compositesa.2008.02.016}}.

\bibitem{Boisse2011}
P.~Boisse, N.~Hamila, E.~Vidal-Salle, F.~Dumont, {Simulation of wrinkling
  during textile composite reinforcement forming. Influence of tensile,
  in-plane shear and bending stiffnesses}, Composites Science and Technology
  71~(5) (2011) 683--692.
\newblock \href {http://dx.doi.org/10.1016/j.compscitech.2011.01.011}
  {\path{doi:10.1016/j.compscitech.2011.01.011}}.

\bibitem{Newell1995}
G.~C. Newell, K.~Khodabandehloo, {Modelling Flexible Sheets for Automatic
  Handling and Lay-up of Composite Components}, Proceedings of the Institution
  of Mechanical Engineers, Part B: Journal of Engineering Manufacture 209~(6)
  (1995) 423--432.
\newblock \href {http://dx.doi.org/10.1243/pime_proc_1995_209_106_02}
  {\path{doi:10.1243/pime_proc_1995_209_106_02}}.

\bibitem{Peng2005}
X.~Peng, J.~Cao, {A continuum mechanics-based non-orthogonal constitutive model
  for woven composite fabrics}, Composites Part A: Applied Science and
  Manufacturing 36~(6) (2005) 859--874.
\newblock \href {http://dx.doi.org/10.1016/j.compositesa.2004.08.008}
  {\path{doi:10.1016/j.compositesa.2004.08.008}}.

\bibitem{Harrison2016}
P.~Harrison, {Modelling the forming mechanics of engineering fabrics using a
  mutually constrained pantographic beam and membrane mesh}, Composites Part A:
  Applied Science and Manufacturing 81 (2016) 145--157.
\newblock \href {http://dx.doi.org/10.1016/j.compositesa.2015.11.005}
  {\path{doi:10.1016/j.compositesa.2015.11.005}}.

\bibitem{Krogh2019}
C.~Krogh, J.~A. Glud, J.~Jakobsen, {Modeling the robotic manipulation of woven
  carbon fiber prepreg plies onto double curved molds: A path-dependent
  problem}, Journal of Composite Materials 53~(15) (2019) 2149--2164.
\newblock \href {http://dx.doi.org/10.1177/0021998318822722}
  {\path{doi:10.1177/0021998318822722}}.

\bibitem{Mack1956}
C.~Mack, H.~M. Taylor, {The Fitting of Woven Cloth to Surfaces}, Journal of the
  Textile Institute Transactions 47~(9) (1956) T477--T488.
\newblock \href {http://dx.doi.org/10.1080/19447027.1956.10750433}
  {\path{doi:10.1080/19447027.1956.10750433}}.

\bibitem{Breen1994}
D.~E. Breen, D.~H. House, M.~J. Wozny, {A Particle-Based Model for Simulating
  the Draping Behavior of Woven Cloth}, Textile Research Journal 64~(11) (1994)
  663--685.
\newblock \href {http://dx.doi.org/10.1177/004051759406401106}
  {\path{doi:10.1177/004051759406401106}}.

\bibitem{Boubaker2006}
B.~{Ben Boubaker}, B.~Haussy, J.~F. Ganghoffer, {Discrete models of woven
  structures. Macroscopic approach}, Composites Part B: Engineering 38~(4)
  (2007) 498--505.
\newblock \href {http://dx.doi.org/10.1016/j.compositesb.2006.01.007}
  {\path{doi:10.1016/j.compositesb.2006.01.007}}.

\bibitem{Do2006}
D.~Do, S.~John, I.~Herszberg, {3D deformation models for the automated
  manufacture of composite components}, Composites Part A: Applied Science and
  Manufacturing 37~(9) (2006) 1377--1389.
\newblock \href {http://dx.doi.org/10.1016/j.compositesa.2005.07.011}
  {\path{doi:10.1016/j.compositesa.2005.07.011}}.

\bibitem{Belnoue2018}
J.~P. Belnoue, O.~J. Nixon-Pearson, A.~J. Thompson, D.~S. Ivanov, K.~D. Potter,
  S.~R. Hallett, {Consolidation-driven defect generation in thick composite
  parts}, Journal of Manufacturing Science and Engineering, Transactions of the
  ASME 140~(7).
\newblock \href {http://dx.doi.org/10.1115/1.4039555}
  {\path{doi:10.1115/1.4039555}}.

\bibitem{Thompson2018}
A.~J. Thompson, J.~P. Belnoue, S.~R. Hallett, {A numerical study examining the
  formation of consolidation induced defects in dry textile composites}, in:
  IOP Conference Series: Materials Science and Engineering, Vol. 406, 2018.
\newblock \href {http://dx.doi.org/10.1088/1757-899X/406/1/012052}
  {\path{doi:10.1088/1757-899X/406/1/012052}}.

\bibitem{Simacek2020}
P.~{\v{S}}im{\'{a}}{\v{c}}ek, S.~G. Advani, {A continuum approach for
  consolidation modeling in composites processing}, Composites Science and
  Technology 186 (2020) 107892.
\newblock \href {http://dx.doi.org/10.1016/j.compscitech.2019.107892}
  {\path{doi:10.1016/j.compscitech.2019.107892}}.

\bibitem{Dodwell2014}
T.~J. Dodwell, R.~Butler, G.~W. Hunt, {Out-of-plane ply wrinkling defects
  during consolidation over an external radius}, Composites Science and
  Technology 105 (2014) 151--159.
\newblock \href {http://dx.doi.org/10.1016/j.compscitech.2014.10.007}
  {\path{doi:10.1016/j.compscitech.2014.10.007}}.

\bibitem{Levy2019}
A.~Levy, P.~Hubert, {Vacuum-Bagged Composite Laminate Forming Processes:
  Predicting Thickness Deviation in Complex Shapes}, Composites Part A: Applied
  Science and Manufacturing 126 (2019) 105568.
\newblock \href {http://dx.doi.org/10.1016/j.compositesa.2019.105568}
  {\path{doi:10.1016/j.compositesa.2019.105568}}.

\bibitem{Hallander2013}
P.~Hallander, M.~Åkermo, C.~Mattei, M.~Petersson, T.~Nyman, {An experimental
  study of mechanisms behind wrinkle development during forming of composite
  laminates}, Composites Part A: Applied Science and Manufacturing 50 (2013)
  54--64.
\newblock \href {http://dx.doi.org/10.1016/j.compositesa.2013.03.013}
  {\path{doi:10.1016/j.compositesa.2013.03.013}}.

\bibitem{Lightfoot2013}
J.~S. Lightfoot, M.~R. Wisnom, K.~Potter, {A new mechanism for the formation of
  ply wrinkles due to shear between plies}, Composites Part A: Applied Science
  and Manufacturing 49 (2013) 139--147.
\newblock \href {http://dx.doi.org/10.1016/j.compositesa.2013.03.002}
  {\path{doi:10.1016/j.compositesa.2013.03.002}}.

\bibitem{Reich1997}
C.~Reich, R.~Ritter, J.~Thesing, {White light heterodyne principle for
  3D-measurement}, in: O.~Loffeld (Ed.), Sensors, Sensor Systems, and Sensor
  Data Processing, Vol. 3100, SPIE, 1997, pp. 236--244.
\newblock \href {http://dx.doi.org/10.1117/12.287750}
  {\path{doi:10.1117/12.287750}}.

\bibitem{matlab}
{The MathWorks, Inc.},
  \href{https://www.mathworks.com/help/matlab/index.html}{{MATLAB
  Documentation}} (2019).
\newline\urlprefix\url{https://www.mathworks.com/help/matlab/index.html}

\bibitem{Rusu2008}
R.~B. Rusu, Z.~C. Marton, N.~Blodow, M.~Dolha, M.~Beetz, {Towards 3D Point
  cloud based object maps for household environments}, Robotics and Autonomous
  Systems 56~(11) (2008) 927--941.
\newblock \href {http://dx.doi.org/10.1016/j.robot.2008.08.005}
  {\path{doi:10.1016/j.robot.2008.08.005}}.

\bibitem{Soille1999}
P.~Soille, {Morphological Image Analysis}, Springer Berlin Heidelberg, 1999.
\newblock \href {http://dx.doi.org/10.1007/978-3-662-03939-7}
  {\path{doi:10.1007/978-3-662-03939-7}}.

\bibitem{gonzalez2004}
R.~C. Gonzalez, R.~E. Woods, S.~L. Eddins, {Digital Image Processing Using
  MATLAB}, Pearson Prentice Hall, 2004.

\bibitem{Gere2008}
J.~M. Gere, B.~J. Goodno, {Mechanics of Materials}, Vol.~7, Cengage Learning,
  Stamford, CT, USA, 2008.

\bibitem{Krogh2019c}
C.~Krogh, K.~D. White, A.~Sabato, J.~A. Sherwood, {Picture-frame testing of
  woven prepreg fabric: An investigation of sample geometry and shear angle
  acquisition}, International Journal of Material Forming (2019) 1--13.~In
  press.
\newblock \href {http://dx.doi.org/10.1007/s12289-019-01499-y}
  {\path{doi:10.1007/s12289-019-01499-y}}.

\bibitem{Krogh2017a}
C.~Krogh, J.~A. Glud, J.~Jakobsen, {Modeling of prepregs during automated
  draping sequences}, in: AIP Conference Proceedings, Vol. 1896, AIP
  Publishing, 2017, p. 030036.
\newblock \href {http://dx.doi.org/10.1063/1.5008023}
  {\path{doi:10.1063/1.5008023}}.

\bibitem{DAgostino2015}
M.~V. D'Agostino, I.~Giorgio, L.~Greco, A.~Madeo, P.~Boisse, {Continuum and
  discrete models for structures including (quasi-) inextensible elasticae with
  a view to the design and modeling of composite reinforcements}, International
  Journal of Solids and Structures 59 (2015) 1--17.
\newblock \href {http://dx.doi.org/10.1016/j.ijsolstr.2014.12.014}
  {\path{doi:10.1016/j.ijsolstr.2014.12.014}}.

\end{thebibliography}

\end{document}